\documentclass[prl,twocolumn,aps,amsmath,amsfonts]{revtex4}   
\usepackage{epsfig,psfrag}
\usepackage{amsmath}
\usepackage{latexsym,exscale}
\usepackage{float}
\usepackage{graphics,subfigure}
\usepackage{color}
\usepackage{lscape}

\begin{document}
\title{Electron correlations and single-particle physics in the Integer Quantum Hall Effect}
\author{Alexander Struck and Bernhard Kramer} \affiliation{I. Institut f\"ur
  Theoretische Physik, Universit\"at Hamburg, Jungiusstra\ss{}e 9, 20355
  Hamburg, Germany} 
\date{\today}
\begin{abstract}
  The compressibility of a two-dimensional electron system with spin in a
  spatially correlated random potential and a quantizing magnetic field is
  investigated. Electron-electron interaction is treated with the Hartree-Fock
  method. Numerical results for the influences of interaction and
  disorder on the compressibility as a function of the particle density and
  the strength of the magnetic field are presented.
  Localization-delocalization transitions associated with highly compressible
  region in the energy spectrum are found at half-integer filling factors.
 Coulomb blockade effects are found near integer fillings in the regions of low
  compressibility. Results are compared with recent experiments.
\end{abstract}

\maketitle

The integer quantization of the Hall conductance of a  two-dimensional electron
system (2DEG) in a strong magnetic field\cite{Klitzing} can be
understood in terms of quantum phase transitions near the centres of the
Landau bands associated with disorder-induced localization-delocalization
transitions of single-electron states in a one parameter scaling model. Neglecting interaction, the
localization length has been found diverging, $\xi\propto|E-E_C|^{-\tilde \nu}$,
with the critical energy $E_{\rm c}$. The universal value of
the critical exponent, $\tilde \nu=2.34\pm 0.04$  \cite{HuckesteinKramer,Huckestein-Review}, is widely
accepted. Peaks in the
magneto-conductance are associated with $E_{\rm c}$. The localized states in the band tails are associated with
zero conductance at zero temperature \cite{Prange-book}.

Despite consistency with several transport
experiments \cite{Wei}, the validity of the one-parameter
scaling model has been controversely discussed, in particular the impact of interaction on the frequency-dependent scaling of the conductivity and the tunneling density of states\cite{ChalkerDaniell,Engel,chan}.

In recent experiments, mesoscopic
conductance fluctuations found in silicon metal-oxide-semiconductor field-effect transistors in dc-transport show regular
patterns which have been interpreted as charging
effects \cite{CobdenBarnesFord}. Patterns associated with Coulomb
blockade in localized states have also been found in measurements of the shift $d\mu/dn$
of the chemical potential $\mu$ with a scanning single-electron transistor probe when changing the particle
density $n$ \cite{Yacoby}. The latter results have been interpreted in a model in which the quantum Hall transition appears as a
result of the strong and complete screening of the disorder potential
near half-integer filling factors. The absence of screening in the
incompressible regions of the energy spectrum leads to localized states that
account for the observed charging effects. In this model, the phase transition
has been interpreted as a percolation transition between incompressible and compressible
regions at certain concentrations of localized charge islands.

We report results of an extensive unrestricted
Hartree-Fock (HF) study of the 2DEG with spin in the presence of long-range
correlated disorder and a perpendicular magnetic field that can
contribute towards more detailed understanding of the experiments. We have
studied the change $d\mu/dn$ of the chemical potential $\mu$ with the particle density $n=N/L^2$ of $N$ particles in a square system of linear size $L$ as a function of $n$
and the magnetic field strength $B$. This quantity is proportional to the inverse compressibility $ \kappa^{-1}\propto  d\mu/dn$. We find interaction-induced enhancement of the
$g$-factor and strong evidence for charging effects in regions near integer filling factors. The shapes of the HF quasiparticle wave functions indicate no significant change in the  
localization behavior at the Fermi level compared to
the non-interacting limit. This is due to quantum corrections which modify the
percolation mechanism \cite{Kramerreview}. Our results are consistent with the recent experiment \cite{Yacoby}.
However, we find that the interaction does not destroy the critical behaviour at the quantum Hall phase transition.

The Hamiltonian of the 2DEG in GaAs is $H^{s}_0+V_{\rm C}$ with
$H_0^{s}=({\bf p}- e {\bf A})^2/({2m^\ast})+sg\mu_B B/2+ 
V_{\rm dis}({\bf r})$
with the vector potential ${\bf A}=(0,Bx,0)$, flux density $B$ and spin $s=\pm
1$. $m^\ast=0.067m_{\rm e}$ is the electron mass,  $\mu_B$ is the Bohr magneton and $g=-0.44$ the electron $g$-factor. The impurity potential is  $V_{\rm dis}({\bf r})=
\sum_{i=1}^{N_{\rm i}} ({V_i}/{\pi d^2})\exp{[{({\bf r}-{\bf r}_i)^2}/{d^2}]}$
with $N_{\rm i}$ the number of scatterers at random positions ${\bf r}_i$ with
random strengths $V_i$, $-V_0<V_i<V_0$. The range $d$ of the impurity
potential is the spatial correlation length of the randomness, $d=0$
corresponds to uncorrelated disorder, $d>l_{B}$
($l_{B}=\sqrt{\hbar/m^\ast\omega_c}$ magnetic length, $\omega_c=eB/m$ cyclotron
frequency) yields a slowly varying potential which is believed to be adeqate
for high mobility samples. The disorder introduces the energy scale
$\Gamma=(N_{\rm i}V_0^2/l_{B}^2L^2)^{1/2}$ ($L^2$ area of the 2DEG). The
Coulomb interaction
$V_{\rm C}({\bf r}-{\bf r'})=
e^2/(4\pi\epsilon\epsilon|{\bf r}-{\bf r'}|)$
introduces an energy scale $\gamma=e^2/4\pi\epsilon\epsilon l_B$ ($e$
elementary charge, dielectric constant  $\epsilon=12.4$ ).  Periodic
boundary conditions are assumed \cite{Ando1,Yoshioka}. Neglecting disorder and
interaction, the Schr\"odinger equation yields the Landau wavefunctions
$|mX\rangle$ ($X=k_jl_B^2$ guiding center coordinate, $k_j=2\pi j/L$ wavenumber)
that are used for the construction of the HF basis.

The HF equation is
\begin{equation}\nonumber\label{eq:hf}
\sum\limits_{b}F^{s}_{ab} C_{b}^{\alpha s}=
E^{\alpha s} C_{a}^{\alpha s}
\end{equation}  
where $\langle mX|\alpha s\rangle=C_{mX}^{\alpha s}\equiv C_{a}^{\alpha s}$
are the expansion coefficients of the HF states $|\alpha s\rangle$
and $E^{\alpha s}$ the energy eigenvalues. The Fock matrix
$F^{s}_{mXm'X'}\equiv 
F^s_{ij}= H^s_{0,ij}+ \sum\limits_a\sum\limits_b \rho_{ab} M_{ijab} - 
\rho^s_{a b} M_{iabj}
$
has to be determined selfconsistently. It contains the interaction matrix
elements
$
M_{ijab}= \frac{\gamma l}{L^2} 
\sum\limits_{q_x,q_y} V(q) \langle i|e^{i{\bf qr}}|j\rangle \langle
a|e^{-i{\bf qr}}|b\rangle
$ 
and the density matrix $\rho_{ij}=\sum_{s}\rho^{s}_{ij}=\sum\limits_{\alpha({\rm occ})}
C^{\alpha s \ast}_{i}C^{\alpha s}_{j}$.

We have performed selfconsistent calculations for $B$ in the range of $1\ldots
6$T and $n=(2\ldots 200)/L^2$ in a square of length $L=30 a$ for $a=10$ nm. The sample size is comparable to the sample size studied in Ref. \onlinecite{Yacoby}. This
yields electron densities $n=(0.22\ldots 22)\,10^{10}$\,cm$^{-2}$. We used
250 impurities with $d=2a$ and $V_0=2$ meV to model a high mobility sample.

To study Coulomb interactions, it is useful to control
its strength relative to the other energy scales. We set
$\gamma=c\gamma_0$; $c=0$ corresponds to the noninteracting system, $c=1$ to the non-screened Coulomb interaction. For each combination of parameters, the self-consistent field and thus total energy and chemical potential are determined. Thus, the self-consistent field can relax with respect to changes in parameters and ground state and the quasiparticle wavefunctions are optimized for given $N$ and $B$. This is sometimes referred to as the "delta-SCF-method"  \cite{onidabagus}: The difference between total energies is calculated successfully in HF approximation, because the response of the $N$-particle system to the addition of another electron or hole is contained in the calculation of the $N\pm 1$-particle many-body wavefunction.

 The number of Landau bands per spin included in the calculation was chosen such that if initially the electron with the highest energy at the highest particle number ($N_{\rm max}=200$) is in Landau band $m_{\rm max}$, $ m_{\rm max}+1$ Landau bands per spin are used in order to provide enough states for the self-consistent field calculations; $N_\phi=L^2/(2\pi l_B^2)=BL^2e/h$ is the number of flux quanta in the system and equals the number of states per spin in a Landau band. For example, at $B=1$ T and $N_\phi=21$, we have used 11 Landau bands per spin, in total 462 states to host 2 to 200 electrons. In the results presented below, we focus on particle numbers $2<N<117$.

\begin{figure}[htbp]
  \includegraphics[angle=-90,width=0.5\textwidth]{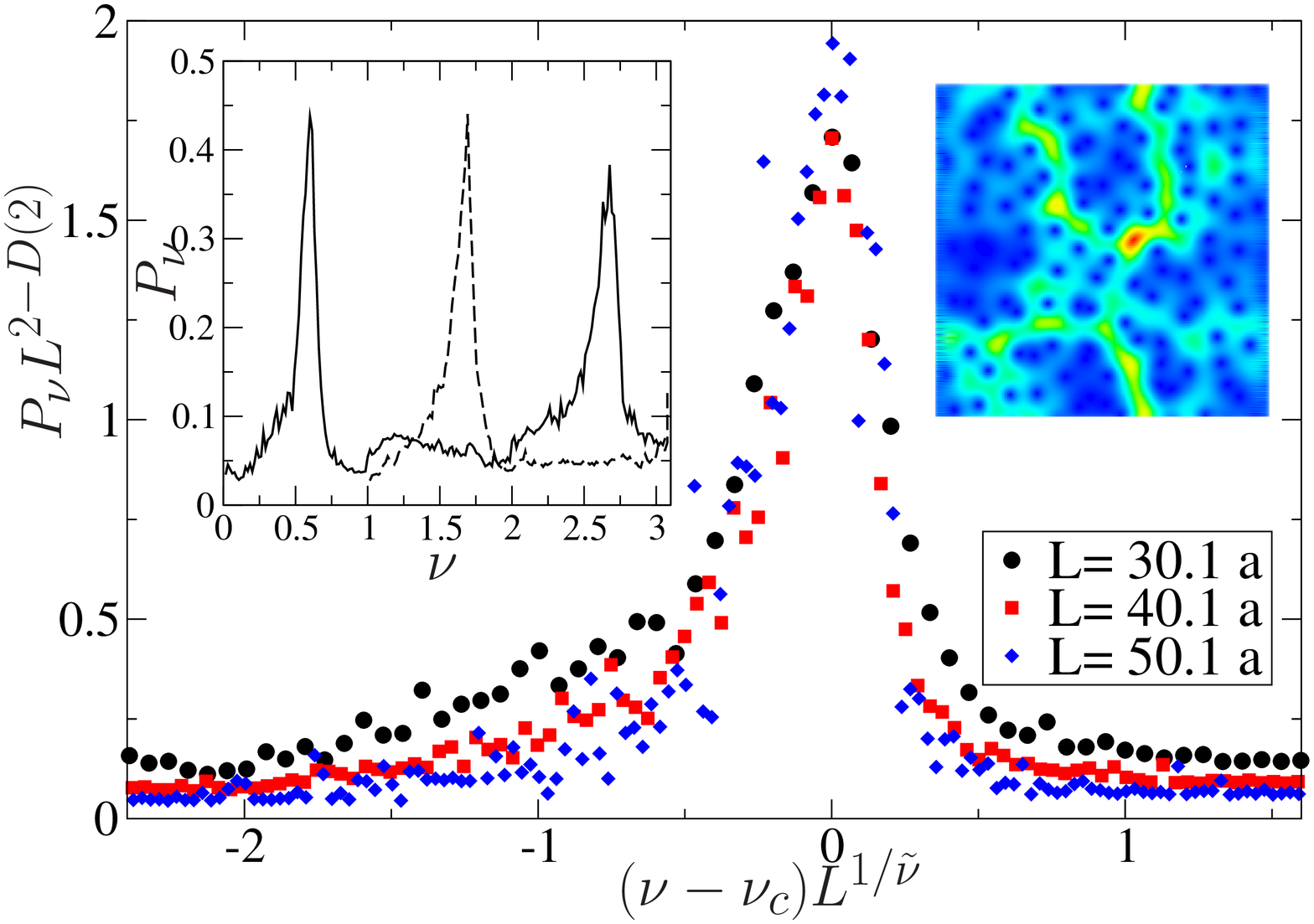}
  \caption{Scaling of the participation ratio $\Pi=L^{2-D(2)}P$ as a function of the filling factor $(\nu-\nu_c)L^{1/\tilde\nu}$, disorder-averaged for each system length $L$ \cite{HuckesteinYangMacDonald}. Parameters 
   $\tilde\nu=2.3$ (critical exponent) and $D(2)=1.6$ (correlation dimensions of wavefunctions \cite{HuckesteinSchweitzer}) 
   Lead to reasonable data collapse.
   Inset: Dependence of $P_\nu(E_F)$ on $\nu$ for spin-up (solid) and spin-down (dashed) for $L=30.1 a$, disorder-averaged. The $g$-factor enhancement is so large, that a subsequent filling of the spin-split Landau level is observed.
   Color inset: Typical probability density [red (bright) high, blue (dark) low value] of a HF state near the band center.}
  \label{fig:struckqhefig1}
\end{figure}

For energies in the band tails, the wave functions are localized near equipotential lines, at least for strong magnetic field. In the center of the band, the wave
functions are delocalized (Fig. \ref{fig:struckqhefig1} inset). 

The scaling of the participation number $P$ at the Fermi energy $E$, $P^{-1}=\int {\rm d}^{2}r|\psi^{4}({\bf
  r})|^{4}\propto \xi^{-2}(E)$ as a function of the filling factor
(Fig.~\ref{fig:struckqhefig1}) has been investigated \cite{HuckesteinYangMacDonald}. With $\tilde \nu=2.3$ resonable collapse to a single curve has been achieved of the data for different system sizes consistent with the critical behavior of the localization length without interaction. The curve resembles the scaling of HF wave functions obtained at \em fixed \rm filling factor using occupied and empty HF orbitals \cite{HuckesteinYangMacDonald}. Within our method the orbitals are separately  determined for every combination of electron density and magnetic field, together with the self-consistent potential. 
The consistency within the errors of the data with the scaling hypothesis indicates that the critical behavior of the HF states is unaffected by the change of the self-consistent field induced by changing the filling factor. However, this must be confirmed by more precise scaling studies for larger systems \cite{HuckesteinKramer}.
The HF energy for $N$ particles
$E_{\rm HF}^{B,N}=\frac{1}{2} \sum\limits_{abs}\rho_{ab}^s\left (H_{0,ab}^s+ F^s_{ab}\right )$
is used to determine the chemical potential $\mu=E_{\rm HF}^{N+1,B}-E_{\rm HF}^{N,B}$ and 
\begin{equation}\nonumber
\frac{d\mu}{dn}=L^2\left ( E_{\rm HF}^{N+1,B}-2E_{\rm HF}^{N,B}+E_{\rm HF}^{N-1,B}\right )\propto \frac{1}{ \kappa}
\label{eq:inv_comp_def}
\end{equation}
This definition reflects a global property, in contrast to the local compressibility reported in Ref. \onlinecite{Yacoby}.
However, the total energy and its derivative depend on the density matrices $\rho^{\uparrow,\downarrow}$ formed with the HF orbitals. The latter are obtained for a finite system and can be considered to reflect local properties such as a specific disorder potential and electrostatic and exchange interaction with the surrounding electrons. Thus, the inverse compressibility calculated here can be expected to reflect the features observed in the measurements done with a tunnel tip.

\begin{figure}[htbp]
\includegraphics[width=0.5\textwidth]{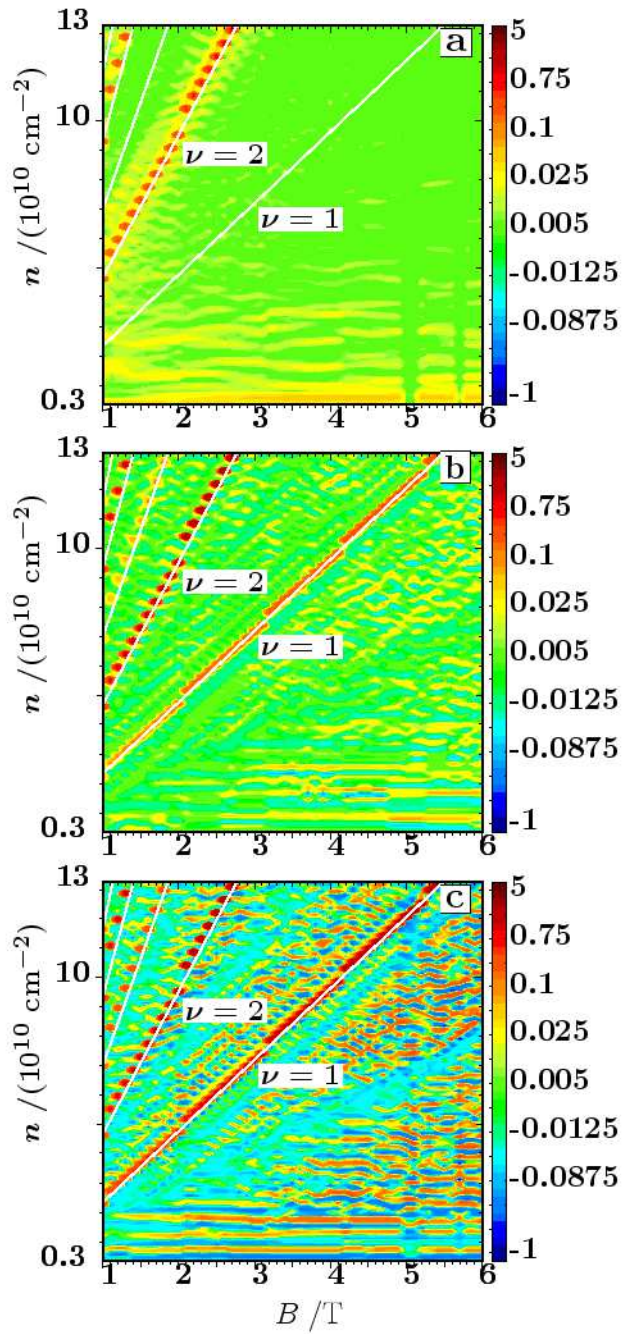}
\caption{$d\mu/dn$ (in units of meV/$L^2$) in the ($n,B$)-plane 
  for interacting electrons (a) $\gamma/\gamma_0=0$, (b) $\gamma/\gamma_0=0.1$,
  (c) $\gamma/\gamma_0=0.5$.  Dark (blue) regions high, bright (red) regions low compressibility. 
  Solid white lines are guide to the eyes.} 
\label{fig:struckqhefig2}
\end{figure} 
Figure~\ref{fig:struckqhefig2} shows $d\mu/dn$ for various particle
numbers and magnetic fields without and with interaction. Without interaction (Fig.  ~\ref{fig:struckqhefig2}a),
states are compressible almost everywhere except near even integer filling, $\nu=N/N_\phi$. This is expected since every particle that enters finds many states at energies close to the Fermi level.
The Fermi energy remains almost constant with respect to the
particle number, as long as the actual Landau band is not completely filled.
Near even integer filling, the Fermi level jumps to the next Landau level due to
very small level density. In the presence of Zeeman
splitting, there are $N_\phi$ states per spin. Since the Zeeman energy in GaAs is very small as compared to
other energy scales, the levels overlap strongly and are equally occupied.
Incompressible lines are then only obtained for even fillings.
With interaction, the Zeeman splitting is large due to exchange enhancement of
the $g$-factor \cite{Ando2}. Spin-up and spin-down levels are
well separated, resulting in additional incompressible lines at odd
fillings. This effect is exaggerated in the unrestricted
HF approximation \cite{MacDonaldOjiLiu}.

Figures  \ref{fig:struckqhefig2}b,c show compressibility patterns for different
interaction strengths.
The compressibility of the interacting electrons shows several regular
structures. Horizontal lines of
constant compressibility parallel to the $B$-axis appear below $\nu=1/2$, enclosed by lines of low or even negative compressibility. At $\nu\approx 1/2$, there is a region of high compressibility. For filling
$\nu\approx 1$, lines of low compressibility parallel to $n=\nu
n_{B}=j/2\pi l_{B}^{2}=jeB/h$ are
observed. The range of electron density where this happens is
independent of $B$. The number of the strongly localized states must
therefore be independent of $B$. This has been ascribed before to Coulomb
blockade in strongly localized states associated with deep potential
wells \cite{Yacoby}. When the potential landscape is completely screened the addition of a further electron is possible. This can even result in a negative compressibility, which in this case is not related to a thermodynamic instability, but to the fact that a positive impurity can be overcompensated by an entering electron. This in turn causes a depletion of electronic charge afterwards in this region \cite{March}.
These charging effects occur also in the higher Landau levels, although less prominent, because 
the localization length of the electrons increases with the Landau level index, thus making the
system more sensitive to compression. We emphasize that the
behavior of the HF total energy with $n$ and $B$ provides the
correct charging pattern in the strongly localized region.

In addition to these Coulomb interaction-dominated features, 
we observe highly compressible regions around half integer
fillings $\nu=j/2$, $n_{B}=jeB/2h$. These correspond to the centers of
the Landau bands where the disorder is considerably screened. The width of
these regions, $\Delta n_{\rm ext}$, is roughly constant as a function of $B$.
The number of the effectively extended states (diameters
larger than $L$) in a Landau band must be
almost independent of $B$ although the total number of single-electron states per
Landau band increases linearly with $B$. This can be understood
 in the one-band approximation, where the single particle density of
states $D$ scales as $D(E/\Gamma)=(N_{\phi}/\Gamma)f(EB/\Gamma)$ \cite{schweitzer}. The
energy interval $\Delta E=|E-E_{\rm c}|$ in which the localization length
exceeds the system size is defined by $\xi(E)=\xi_{0}|E-E_{0}|^{-\tilde \nu}>L$.
Thus, $\Delta E = (L/\xi_{0})^{-1/\tilde \nu}\propto B^{-1/\tilde \nu}$, since
$\xi_{0}\propto \Gamma^{-2}\propto B^{-1}$, and $\delta n_{\rm ext}\approx
D(0)\Delta E\propto B^{1/2-1/\tilde\nu}=B^{0.065}$ with $\nu\approx
2.3$. 

In contrast to previous assertions \cite{Yacoby}, we find between the low-compressibility regions of Coulomb blockade in the strongly
localized states and the high compressibility regions of delocalization, that there
are large regions of intermediate statistically fluctuating compressibility.
These correspond to localized states that cover larger spatial regions with
randomly fluctuating areas. The charging energies of these states, if
applicable at all, should be much smaller than in the regimes of strongly
localized states, and also strongly fluctuating. As a consequence, one would
not expect regular compressibility patterns in these intermediate regions, and
this is what is observed in Fig.~\ref{fig:struckqhefig2}c. In these regions,
the localization properties are determined by the competition of tunneling
{\em between}, and destructive interference {\em along} the percolating
equipotential lines, and it is this competition that is responsible for the
critical behavior \cite{Kramerreview}. The regimes of strong localization and  extended states are clearly observed, separated by regions of intermediate states. 

This could be affected by the finite system size $L$, if the localization length of the intermediate states is comparable to $L$. Here, it is likely that the fluctuations vanish in the thermodynamic limit, and only incompressible localized stripes remain, consistent with experiment. Preliminary results for larger systems show that the fluctuations seem to remain present. A systematic study of the size dependence will be reported elsewhere \cite{struck-unpublished}.

In conclusion, we have investigated the density dependence of the chemical potential as a function of electron density and magnetic field for a quantum Hall system. We have shown that electron interactions, treated in HF approximation, but with the possibility for the ground state to respond to changes in magnetic field or electron density, modify the compressibility pattern. The appearance of regular structures can be interpreted as charging of localized states. This is consistent with recent experiments and suggests that interactions are important for the understanding of the integer quantum Hall effect, especially in the plateau regions. However, the results reported here are not in contradiction to the conjecture that the critical behavior of the metal-insulator transition is unaffected by interactions and microscopic details of the disorder potential since the scaling of the participation number at the Fermi energy is found to be consistent with the scaling hypothesis, although for each electron density the effective potential changes as a result of charge rearrangement.
  
The question remains open if correlations beyond the HF approximation can affect the compressibility pattern. Calculations regarding correlation effects have not been done. However, the HF results reproduce the charging effects in the regimes of localized states well and support the assumption that the critical behaviour of the integer quantum Hall transition can be understood within a single particle model.

This work has been supported by the EU grant MCRTN-CT2003-504574.
We thank J. Chalker and S. Kettemann for useful discussions.

\nopagebreak

\end{document}